\documentclass[reprint,
nofootinbib,
 amsmath,amssymb,
 aps,
]{revtex4-1}

\usepackage{amsmath}
\usepackage{graphicx,color}
\usepackage{dcolumn}
\usepackage{bm}

\newcommand{\cb}{{\bar{c}}}

\begin{document}


\title{One-loop unquenched three-gluon and ghost-gluon vertex in the Curci-Ferrari model}

\author{Felipe Figueroa}
\affiliation{Laboratoire d’Annecy-le-Vieux de Physique Th\'eorique LAPTh, Universit\'e Savoie Mont Blanc, CNRS, F-74000 Annecy, France}%

\author{Marcela Pel\'aez}
\affiliation{Instituto de F\'isica, Facultad de Ingenier\'ia, Universidad de la Rep\'ublica, Montevideo, Uruguay}%

\date{\today}

\begin{abstract}
In this article we study the unquenched three-gluon and ghost-gluon vertex in all momentum range going from the ultraviolet to the infrared regime using the Curci-Ferrari model at one-loop in Landau gauge as an extension of the results presented in \cite{Pelaez:2013cpa}. 
Results are compared with recent lattice data for $SU(3)$ in the unquenched case. 
This calculation is a pure prediction of the model because it does not require fixing any parameter once two-point functions are fitted. 
An analysis of the influence of dynamical quarks in the position of the zero crossing of the three-gluon vertex is presented.
Due to the recent improvements in infrared lattice data for the quenched three-gluon correlation function \cite{Aguilar:2021lke} we also redo the comparison between our one-loop 
results in this limit and the lattice obtaining a very good match.
\end{abstract}

\maketitle


\section{\label{sec_intro}Introduction}

The infrared sector of QCD is usually called the Nonperturbative regime due to the fact that standard perturbation theory based on Faddeev-Popov Lagrangian presents a Landau pole in the infrared. 
This implies that perturbation theory cannot be applied together with this particular gauge-fixed Lagrangian to study the low energy region. 
For these reasons different semi-analytical alternatives have been developed in order to approach this regime. 
For instance, approaches using non-perturbative functional techniques as treatments based on Schwinger-Dyson equations (SD) 
(see e.g. \cite{Alkofer:2000wg, vonSmekal:1997ohs,Atkinson:1998zc,Zwanziger:2001kw,Lerche:2002ep,Fischer:2002hna,Maas:2004se,Boucaud:2006if,Huber:2007kc,Aguilar:2007jj,Aguilar:2008xm,Boucaud:2008ji,Boucaud:2008ji,Boucaud:2011ug,DallOlio:2012own,Huber:2012zj,Aguilar:2013xqa,Aguilar:2015nqa,Huber:2016tvc,Papavassiliou:2017qlq,Huber:2020keu,Aguilar:2020yni,Gao:2021wun}), 
the functional renormalization group \cite{Ellwanger:1996wy,Pawlowski:2003hq,Fischer:2004uk,Fischer:2006vf,Fischer:2008uz,Cyrol:2016tym,Dupuis:2020fhh} 
or the variational Hamiltonian approach \cite{Schleifenbaum:2006bq}. 
Other approaches focus on 
the fact that Faddeev-Popov procedure, generally used to fixed the gauge, is not completely justified in the infrared 
due to the existence of Gribov copies \cite{Gribov:1977wm}. However, until now, it is not known how to 
build a gauge-fixing procedure from first principles taking into account properly the problem of Gribov copies in the infrared.
There are some interesting attempts to reach this gauge-fixed Lagrangian based on Gribov-Zwanziger action and the refined Gribov-Zwanziger 
approach \cite{Zwanziger:1989mf,Vandersickel:2012tz,Dudal:2008sp}.

On top of these semi-analytical studies there are lattice simulations.
Lattice simulations can deal with the problem of Gribov copies so they are a good tool to obtain information about the infrared behavior of Yang-Mills theory. 
Two important observations of lattice simulations are, first, that the gluon propagator 
reaches a finite nonzero value in the infrared, thus behaving as a massive-like propagator in this region. 
\cite{Cucchieri:2007rg,Bogolubsky:2009dc,Bornyakov:2009ug,Iritani:2009mp,Maas:2011se,Oliveira:2012eh}. 
Second, that the relevant expansion parameter obtained through the ghost-gluon or the three-gluon vertex does not present a Landau-pole and in fact it does not become 
too large \cite{Bogolubsky:2009dc,Boucaud:2011ug,Boucaud:2018xup,Zafeiropoulos:2019flq}. 
These points have motivated us to study the infrared regime using a gauge-fixed Lagrangian with a gluon mass term \cite{Tissier:2010ts,Tissier:2011ey}. 
This Lagrangian is a particular case of Curci-Ferrari Lagrangians in Landau gauge (CF) \cite{Curci:1976bt}. 
On a different approach, the mentioned lattice results also motivates a screened massive perturbation theory where 
the mass term is added and subtracted changing the starting theory for expansion 
\cite{Siringo:2015wtx,Siringo:2016jrc,Siringo:2018uho,Comitini:2020ozt}.

Even though we do not know how to justify the CF Lagrangian from first principles it is important to observe that it can reproduce a great variety of 
correlation functions using the first order in perturbation theory. It is important to mention that we do not attempt to reproduce all infrared quantities of QCD perturbatively. In particular the perturbative expansion for correlation functions involving quarks 
within CF near the chiral limit fails. Other approach using CF model was proposed in \cite{Pelaez:2017bhh,Pelaez:2020ups} in order to explore the chiral limit. 
See \cite{Pelaez:2021tpq} for a detail summary of the already studied properties of the model. In particular, one-loop corrections within the CF model were computed for 
propagators, ghost-gluon vertex and the quenched three-gluon vertex \cite{Tissier:2011ey,Pelaez:2013cpa,Pelaez:2014mxa,Pelaez:2015tba,Reinosa:2017qtf}. In addition to this, two-loop corrections were studied for 
propagators \cite{Gracey:2019xom,Barrios:2021cks} and the 
ghost-gluon vertex with a vanishing gluon momentum \cite{Barrios:2020ubx} and compared with lattice data with great accuracy.
It is important to mention that vertices are obtained as a pure prediction of the model, in the sense that the free parameters are fixed by minimizing the error between propagators
and the corresponding lattice data 
and therefore there are no free parameters left when studying vertices. 

The aim of this article is to extend the study of one-loop corrections for the three gluon vertex in the presence of dynamical quarks. 
The infrared regime of the three-gluon vertex has been studied by different 
approaches (see e.g. \cite{Alkofer:2004it,Cucchieri:2006tf,Cucchieri:2008qm,Huber:2012zj,Pelaez:2013cpa,Aguilar:2013vaa,Blum:2014gna,Eichmann:2014xya,Mitter:2014wpa,Williams:2015cvx,Blum:2015lsa,Blum:2016fib,Cyrol:2016tym,Athenodorou:2016oyh,Duarte:2016ieu,Corell:2018yil,Boucaud:2017obn,Binosi:2017rwj,Vujinovic:2018nqc,Aguilar:2019jsj,Aguilar:2019uob,Aguilar:2019kxz,Souza:2019ylx,Aguilar:2021lke}) as it is an important ingredient to understand QCD at low energies. 
The three-gluon vertex is more difficult to calculate than the propagators because instead of depending on a single momentum, 
it depends on three independent scalars. Moreover, it has a richer tensorial decomposition so different scalar functions (associated with different tensors) 
have to be reproduced together.

In this article, we study the one-loop effects of dynamical quarks in the three-gluon vertex using the CF model. The unquenched results are compared with lattice data from \cite{Sternbeck:2017ntv}. 
Moreover, recent simulations of the quenched three-gluon vertex show a better handling of the infrared regime, yielding more precise data in this limit \cite{Aguilar:2021lke}. 
For this reason it is worth to extend the results presented in \cite{Pelaez:2013cpa} for $SU(2)$ to $SU(3)$ gauge-group and compare it with the newest lattice data. For both cases, 
quenched and unquenched, the parameters used in the plots were chosen to minimize the error (understood as discrepancy with the lattice) of the propagators previously computed in \cite{Pelaez:2014mxa,Gracey:2019xom}. 
In this sense, the results shown in this article are a pure prediction of the model that reproduces with great accuracy the lattice data. Due to the presence of massless ghosts, 
CF model also features a zero crossing as it is observed in 
\cite{Pelaez:2013cpa,Aguilar:2013vaa,Blum:2014gna,Eichmann:2014xya,Williams:2015cvx,Blum:2015lsa,Blum:2016fib,Athenodorou:2016oyh,Cyrol:2016tym,Duarte:2016ieu, Boucaud:2017obn,Aguilar:2019jsj,Aguilar:2019uob,Aguilar:2019kxz, Huber:2020keu, Aguilar:2021lke}. 
We also find that the inclusion of dynamical quarks shifts the zero crossing towards the infrared in a way consistent with what is observed in \cite{Williams:2015cvx,Blum:2016fib}.

The article is organized as follows. In Sec. II
we describe in more detail the Curci-Ferrari model in Landau gauge. 
We give some details on the one-loop calculations of the three-gluon vertex in Sec. III in terms of the Ball-Chiu components. 
In Sec. IV we present the renormalization conditions and the renormalization group equations. 
We present our results in Sec. V. and compare them with lattice data. At the end of the article we present the conclusions of the results.

\section{Curci-Ferrari model with quarks}
\label{sec_CF}

We start by introducing the Curci-Ferrari Lagrangian \cite{Curci:1976bt} in the presence of dynamical quarks in Euclidean space:

\begin{equation}
  \label{eq_lagrang}
  \begin{split}
  \mathcal{L}= \frac{1}{4}(F_{\mu\nu}^a)^2+\partial _\mu\overline c^a(D_\mu
  c)^a+ih^a\partial_\mu A_\mu^a \\ +\frac{m^2}{2} (A_\mu^a)^2+\sum_{i=1}^{N_f} \bar{\psi}_i (\gamma_\mu D_\mu + M_i)\psi_i \,,
    \end{split}
\end{equation}
where $g$ is the coupling constant, and the flavor index $i$ spans the $N_f$ quark flavors.

The covariant derivative $D_\mu$ acting on a ghost field in the adjoint representation of SU(N) reads $(D_\mu c)^a=\partial_\mu c^a+g f^{abc}A_\mu^b c^c$, while when applied to a quark in the fundamental 
representation it reads $D_\mu \psi=\partial_\mu \psi-g t^{a}A_\mu^a \psi$. The latin indices correspond to the $SU(N)$ gauge-group, the $t^a$ are the generators in the fundamental representation and the $f^{abc}$ 
are the structure constants of the group. Finally, the field strength is given by $F_{\mu\nu}^a=\partial_\mu A_\nu^a-\partial_\nu A_\mu^a+gf^{abc}A_\mu^bA_\nu^c$.

The Feynman rules associated to this Lagrangian are the standard Feynman rules for Euclidean-QCD in Landau gauge except for the gluon's free propagator, which reads

\begin{equation}
  \label{eq_propag_AA}
  \langle
  A_\mu^aA_\nu^b\rangle_0(p)=\delta^{ab} P^\perp_{\mu\nu}(p)\frac 1{p^2+m^2} \,,
\end{equation}
where we have introduced the transverse projector:
\begin{equation}
  P^\perp_{\mu\nu}(p)=\delta_{\mu\nu}-\frac{p_\mu p_\nu}{p^2}.
\end{equation}

It is important to mention that the gluon mass term added to the Faddeev-Popov Lagrangian breaks the BRST symmetry. However, it can be shown that (\ref{eq_lagrang}) satisfies a modified (non-nilpotent) BRST symmetry that can be used to prove its renormalizability \cite{deBoer:1995dh}. 

Probably the most interesting aspect of this model is the fact that, as it has been shown in various previous articles (see \cite{Tissier:2010ts,Tissier:2011ey,Weber:2014lxa,Reinosa:2017qtf,Gracey:2019xom} 
for instance), the addition of a gluon mass term regularizes the theory in the infrared, allowing for a perturbative treatment of the theory in this region. More specifically, 
it is possible to find a renormalization scheme without an infrared Landau pole for particular choices of the initial condition of the renormalization-group flow. 
These features have made it possible to use this model to compute various two and three-point functions to 1-loop and 2-loop order, obtaining a very good match with lattice simulations 
\cite{Tissier:2011ey,Pelaez:2013cpa,Pelaez:2014mxa,Pelaez:2015tba,Gracey:2019xom,Barrios:2020ubx,Barrios:2021cks}.
It is important to mention that this model has also been studied at finite temperature and chemical potential in \cite{Serreau:2017lxd,Maelger:2017amh,Maelger:2018vow}.

\section{One-loop calculation of the three gluon vertex}

\subsection{Tensorial structure and computation}

In this work we extend the one-loop computation of the gluon three-point function obtained in \cite{Pelaez:2013cpa} for $SU(N)$ Yang-Mills theory to unquenched QCD.
In order to calculate the gluon's three-point function at one-loop order we need to compute the Feynman diagrams shown in Fig \ref{fig_diagramas}.

\begin{figure}[htbp]
 \includegraphics[width=0.9\linewidth]{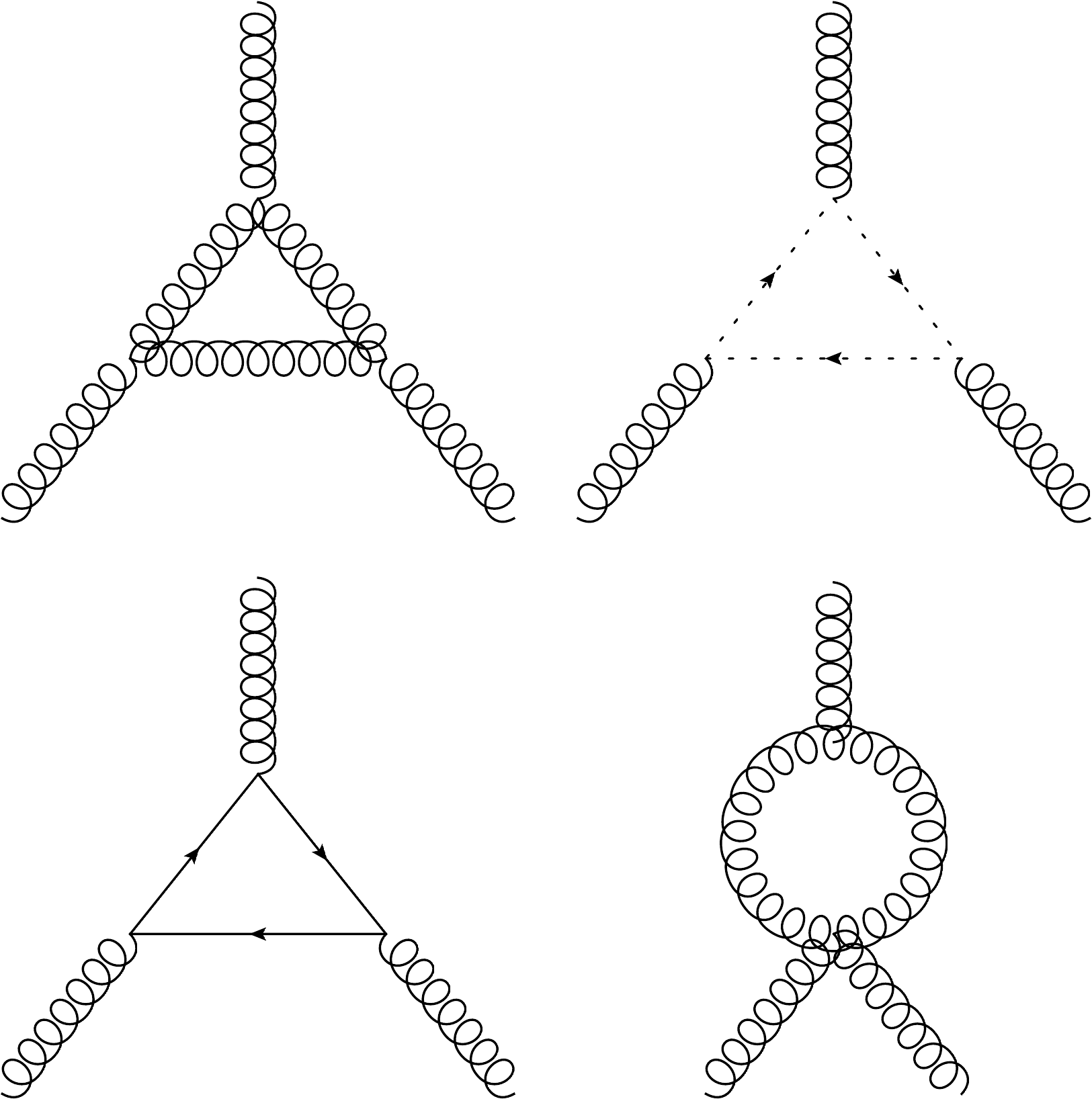}
\caption{Feynman diagrams present in the one-loop calculation of the three-gluon vertex.}
\label{fig_diagramas}
 \end{figure}

As shown in \cite{Smolyakov82}, the color structure of the three-gluon vertex is simply the structure constant $f^{abc}$ of the $SU(N)$ group, so it can be factored out.
Furthermore, we follow the usual convention \cite{Davydychev:2001uj} of factorizing the coupling constant, and thus we define:

$$\Gamma_{A^a_\mu A^b_\nu A^c_\rho}^{(3)}(p,k,r)=-ig f^{abc}\Gamma_{\mu\nu\rho}(p,k,r)$$.

The tensor structure of $\Gamma_{\mu\nu\rho}(p,k,r)$ can be easily deduced: it must depend on three Lorentz indices (one for each gluon) and on two independent momenta due to momentum conservation. 
As a consequence, we can have only two types of tensor structures: the ones made up of three momenta ($p_{\mu} p_{\nu} p_{\rho}$, $p_{\mu} p_{\nu} k_{\rho}$,...) and the ones made up of one momentum and 
the Euclidean metric tensor ($p_{\mu} \delta_{\nu \rho}$, $k_{\nu} \delta_{\mu \rho}$,...).
It's not hard to convince oneself that there are eight possible terms of the first kind and six of the second, adding up to a total of 14 possible terms in the vertex's tensor structure. 
However, the vertex is symmetric under the exchange of any pair of external legs, and this ends up reducing the total number of possible independent coefficients in the vertex's tensor structure to six.

A cleaner way of exploiting these symmetries is following the decomposition proposed by Ball and Chiu in \cite{BallChiu}, where they parametrize the vertex using six scalar functions:

\begin{widetext}
\begin{equation}
  \begin{split}
\Gamma_{\mu\nu\rho}(p,k,r)&=A(p^2,k^2,r^2)\delta_{\mu\nu}(p-k)_\rho+ 
B(p^2,k^2,r^2)\delta_{\mu\nu}(p+k)_\rho 
-C(p^2,k^2,r^2)(\delta_{\mu\nu}p.k-p_{\nu}k_{\mu})(p-k)_\rho\\
&+\frac{1}{3}S(p^2,k^2,r^2)(p_{\rho}k_{\mu}r_{\nu}+p_{\nu}k_{\rho}r_{\mu})+  
F(p^2,k^2,r^2)(\delta_{\mu\nu}p.k-p_{\nu}k_{\mu})(p_{\rho}k.r-k_{\rho}p.r)\\
&+H(p^2,k^2,r^2)\left[-\delta_{\mu\nu}(p_{\rho}k.r-k_{\rho}p.r)+\frac{1}{3}(p_{
\rho}k_{\mu}r_{\nu}-p_{\nu}k_{\rho}r_{\mu})\right]+\text{cyclic permutations}  
  \end{split}
\end{equation}
\end{widetext}

The scalar functions have the following symmetry properties: $A$, $C$
and $F$ are symmetric under permutation of the first two arguments;
$B$ is antisymmetric under permutation of the first two arguments; $H$
is completely symmetric and $S$ is completely antisymmetric. 

It is important to note that only some of these scalar functions are accessible through lattice simulations, since they have access to the vertex function only through the correlation function, 
i.e. the vertex contracted with the full external propagators. Since the propagators in Landau gauge are transverse, the longitudinal part of the vertex function is lost in the process when requiring the conservation of momentum at the vertex. In particular 
this means that the $B$ and $S$ functions are not accessible through lattice computations. 

We decomposed every diagram contributing to the three gluon vertex into the Ball-Chiu tensorial structure. In this way we obtained the contribution of each diagram to each of the scalar functions 
$A, B, C, F, H$ and $S$. To perform our computations we expressed the integrals in Feynman diagrams of Fig.~\ref{fig_diagramas} in terms of only three Master Integrals defined following the convention on \cite{Martin:2005qm} as:
\begin{widetext}
\begin{eqnarray}
\label{masters}
& \textbf{A}[m_{1}]= \bar{C} \int d^{d} q \frac{1}{[q^2+m_{1}^{2}]}\nonumber \\ & \textbf{B}[p_{1},m_{1},m_{2}]= \bar{C} \int d^{d} q \frac{1}{[q^2+m_{1}^{2}][(q+p_{1})^2+m_{2}^{2}]} \nonumber \\ & \textbf{C}[p_{1},p_2,m_{1},m_{2},m_3]= \bar{C} \int d^{d} q \frac{1}{[q^2+m_{1}^{2}][(q+p_{1})^2+m_{2}^{2}][(q-p_{2})^2+m_{3}^{2}]}
\end{eqnarray}
\end{widetext}
where $\bar{C}=16 \pi^2 \frac{\bar{\mu}^{2 \epsilon}}{(2 \pi)^d}$, and the regularization scale $\bar{\mu}$ is related to the renormalization scale $\mu$ by $\mu^2=4 \pi e^{-\gamma}\bar{\mu}^2$. The $\textbf{A}$ and $\textbf{B}$- Master Integrals can be solved analytically in $d=4-2\epsilon$ in terms of the external momentum and the masses, but the $\textbf{C}$-Master Integral must be treated numerically except for particular kinemetics. We chose the FIRE5 algorithm \cite{Smirnov:2008iw} to perform the Master Integral reduction, thus obtaining analytic expressions for each of the scalar functions in terms of the three Master 
Integrals for arbitrary momentum configurations. 
The expressions are complicated and not very enlightening, however, the explicit expressions appear in the suplemental material of \cite{Pelaez:2013cpa} for the quenched case while the quark contribution is written in \cite{Davydychev:2001uj}. In the case of one vanishing external momentum the computation becomes much simpler and the result for the quenched vertex function in this 
configuration is given in \cite{Pelaez:2013cpa}, while the 
unquenched case is presented in the Appendix \ref{Ap_UQ}. 

\subsection{Checks}

Various checks for the Yang-Mills part of the result for the three-gluon vertex function had already been made in \cite{Pelaez:2013cpa}. We only need to check the quark triangle diagram to test our unquenched results.
To do this we compared our results to those of \cite{Davydychev:2001uj}, verifying that they yield the same expressions when properly transformed to the Euclidean space. This was expected, 
as the quark triangle diagram is independent from the mass of the gluons and therefore its contribution in the Curci-Ferrari model is the same as in standard QCD.

\section{Renormalization and renormalization group}

In this section we introduce the renormalization scheme that we used in this work and we explain how we implemented renormalization-group ideas to improve our perturbative calculation.

\subsection{Renormalization}

To take care of the divergences appearing in the 1-loop quantities we took the usual approach of redefining the fields, masses and coupling constants through renormalization factors that absorb the infinities. 
The renormalized quantities are defined in terms of the bare ones (denoted with a "0" subscript) as follows:

\begin{align}
 A_0^{a\,\mu}&= \sqrt{Z_A} A^{a\,\mu},\hspace{.5cm}  \psi_0= \sqrt{Z_\psi} \psi,\hspace{.2cm} \nonumber\\
 c_0^{a}&= \sqrt{Z_c} c^{a},\hspace{.5cm}
 \bar c_0^{a}= \sqrt{Z_c} \bar c^{a},\hspace{.2cm} \nonumber\\
g_0&= Z_g g \hspace{.5cm} m_0^2= Z_{m^2} m^2\hspace{.5cm} M_0= Z_{M} M
\end{align}

From now on, all expressions will refer to renormalized quantities unless explicitly stated otherwise. The renormalized propagators and the three-gluon 1PI correlation function are thus defined as:

\begin{align}
\label{renormvertices}
 \Gamma^{(2)}_{A_\mu^aA_\nu^b}(p)&=Z_A \Gamma^{(2)}_{A_\mu^aA_\nu^b,0}(p) \nonumber\\
 \Gamma^{(2)}_{c^a \cb^b}(p)&=Z_c  \Gamma^{(2)}_{c^a \cb^b,0}(p) \nonumber\\
  \Gamma^{(2)}_{\psi \bar\psi}(p)&=Z_\psi   \Gamma^{(2)}_{\psi \bar\psi,0}(p) \nonumber\\
 \Gamma^{(3)}_{A_\mu^aA_\nu^bA_\rho^c}(p,r)&=Z_A^{3/2} \Gamma^{(3)}_{A_\mu^aA_\nu^bA_\rho^c,0}(p,r) 
 \end{align}

\subsection{Infrared Safe renormalization scheme}

To fix the renormalization factors we used the Infrared Safe (IS) renormalization scheme defined in \cite{Tissier:2011ey}. 
It is based on a non-renormalization theorem for the gluon mass \cite{Dudal:2003pe, Wschebor:2007vh, Tissier:2008nw}, and is defined by

\begin{align}
\label{rencond}
&\Gamma^\perp(p=\mu)=m^2+\mu^2, \hspace{.4cm} J(p=\mu)=1,\nonumber\\
&Z_{m^2} Z_A Z_c=1.
\end{align}
where $\Gamma^\perp(p)$ is the transversal part of $ \Gamma^{(2)}_{A_\mu^aA_\nu^b}(p)$ and $J(p)$ is the ghost dressing function. 
To fix the renormalization of the coupling constant we used the Taylor scheme, in which the coupling constant is defined as the ghost-gluon vertex with a vanishing ghost momentum. 
Requiring that the renormalized vertex is finite leads to a relation among the renormalization factors $Z_A$, $Z_c$ and $Z_g$:

\begin{equation}
  \label{eq_taylor}
  Z_g\sqrt{Z_A} Z_c=1
\end{equation}
The divergent part of the renormalization factors for the quenched case were presented in \cite{Tissier:2011ey}. Here we show the extension to the unquenched Curci-Ferrari model already computed in \cite{Gracey:2002yt,Pelaez:2014mxa}. In $d=4-2 \epsilon$ they read:
\begin{align}
\label{factoresrenorm}
 &Z_c=1+\frac{3 g^2 N}{64 \pi^2 \epsilon} \nonumber\\
 &Z_A=1+\frac{g^2}{96 \pi^2}\frac{(13N-8 N_f T_f)}{\epsilon} \nonumber\\
  &Z_{m^2}=1-\frac{g^2}{192 \pi^2}\frac{(35N-16 N_f T_f)}{\epsilon} \nonumber\\  &Z_{g}=1-\frac{g^2}{96 \pi^2}\frac{(11N-4 N_f T_f)}{\epsilon}
 \end{align}

Finally, the quantity we are interested in is actually $\Gamma_{\mu\nu\rho}$ as defined earlier. Since in it's definition we factorized a factor of $g$, 
the relation between the renormalized and bare quantities is the following:

$$\Gamma_{\mu\nu\rho}(p,r)=Z_A^{3/2} Z_g \Gamma_{\mu\nu\rho,0}(p,r)=\frac{Z_A}{Z_c}\Gamma_{\mu\nu\rho,0}(p,r),$$
where in the last equality we used equation \ref{eq_taylor}.

\subsection{Renormalization Group}

After the renormalization procedure we obtain a finite expression for the three-gluon vertex, but it comes with the usual loop corrections of the form log$(\frac{p}{\mu})$. 
To handle this situation we implemented the renormalization-group flow to take care of the large logarithms coming from loop corrections. First we define the $\beta$ functions and anomalous dimensions of the fields:

\begin{align}
\beta_\chi(g,m^2,\{M_i\})&=\mu\frac{d\chi}{d\mu}\Big|_{g_0, m^2_0, M_{i,0}},\\
\gamma_\phi(g,m^2,\{M_i\})&=\mu\frac{d\log Z_\phi}{d\mu}\Big|_{g_0, m^2_0, M_{i,0}}.
\end{align}
where $\chi$ can take the role of the coupling constant, the gluon mass or the quark mass and $\phi$ represents the different fields $A,c,\psi$.

The renormalization group equation for the vertex function with $n_A$ gluon legs and $n_c$ ghost legs 
reads:

\begin{equation}
\label{eq_RGequation}
\begin{split}
\Big( \mu \partial_\mu -\frac 1 2 &(n_A \gamma_A+n_c \gamma_c)\\&+\beta_g 
\partial_{g}+
\beta_{m^2}\partial_{m^2}\Big)\Gamma^{(n_A,n_c)}=0,
\end{split}
\end{equation}
This equation allows us to obtain a relation for the vertex function renormalized at scale 
$\mu_0$ and the same vertex function renormalized at a different scale $\mu$:

\begin{equation}
\label{eq_int_RG}
\begin{split}
\Gamma^{(n_A,n_c)}(&\{p_i\};\mu,g(\mu),m^2(\mu),M(\mu))=z_A(\mu)^{
n_A/2}z_c(\mu)^{
n_c/2}\\
&\times\Gamma^{(n_A,n_c)}(\{p_i\};\mu_0,g(\mu_0),m^2(\mu_0),M(\mu_0)).
\end{split} 
\end{equation} 
where $g(\mu)$, $m^2(\mu)$ and $M(\mu)$ are obtained by integration of the
$\beta$ functions with initial conditions given at some scale $\mu_0$ and $z_A$ and $z_c$ are obtained repectively from:  
\begin{equation}
\label{eq_def_z_phi}
\begin{split}
\log z_A(\mu,\mu_0)&=\int_{\mu_0}^\mu\frac
     {d\mu'}{\mu'}\gamma_A\left(g(\mu'),m^2(\mu')\right),\\ \log
     z_c(\mu,\mu_0)&=\int_{\mu_0}^\mu\frac
     {d\mu'}{\mu'}\gamma_c\left(g(\mu'),m^2(\mu')\right).
\end{split}
\end{equation}

We can then use the non-renormalization theorems of Eq.(\ref{rencond})  and Eq.(\ref{eq_taylor}) to relate the anomalous dimensions and the $\beta$ functions. 
It is simple to check that one obtains the following relations:

\begin{align}
\label{eqgama}\gamma_A(g,m^2)&= 2\left(\frac{\beta_{m^2}}{m^2}-\frac{\beta_g}{g}\right),\\
\label{eqgamc}\gamma_c(g,m^2)&=\frac{2\beta_g}{g}-\frac{\beta_{m^2}}{m^2}.
\end{align}

Finally we use these relations to integrate Eq.(\ref{eq_def_z_phi}), obtaining analytical expressions for $z_A$ and $z_c$ in terms of the running gluon mass and coupling constant, 
being this feature another of the advantages of the Infrared Safe scheme:

\begin{equation}
\label{eq_scalefactors}
\begin{split}
 z_A(\mu,\mu_0)&=\frac{m^{4}(\mu)g^{2}(\mu_0)}{m^{4}(\mu_0)g^{2}(\mu)},\\ z_c(\mu,\mu_0)&=\frac{m^{2}(\mu_0)g^{2}(\mu)}{m^{2}(\mu)g^{2}(\mu_0)}.
\end{split}
\end{equation}

We are able now to express the three-gluon vertex renormalized at scale $\mu_0$ in terms of the same quantity using a running scale $\mu= p$, 
thus avoiding the large logarithms problem. Taking into account the factor of g on the definition of $\Gamma_{\mu\nu\rho}(p,r)$ this reads:

\begin{equation}
\label{eq_scalinggamma}
\Gamma_{\mu\nu\rho}(p,r;\mu_0)=\frac{g^{4}(p)m^{6}(\mu_0)}{g^{4}(\mu_0)m^{6}(p)}\Gamma_{\mu\nu\rho}(p,r;\mu=p)
\end{equation}

\section{Results}
We now present our results for the different scalar functions associated to the three-gluon vertex introduced in the previous section. 
All our results correspond to four dimensions and the $SU(3)$ gauge group, and we evaluate the scalar functions in different momentum configurations in order to compare them with the available lattice data.

\subsection{Fixing Parameters}

The only fitting parameters we need to adjust to compare our results with the lattice are the overall normalization constant of the gluon three-point function and the initial conditions of the 
renormalization-group flow, i.e. the values of the mass of the gluon, the mass of the quark and the coupling constant at some renormalization scale $\mu_0$.

The initial conditions for the renormalization-group are best obtained by looking for the set of parameters ($m_0$, $M_0$, $g_0$) that produce the best fit between the gluon and ghost propagators 
computed using the Curci-Ferrari approach and the lattice data, since the lattice results are much more precise for propagators than for three-point functions. This task was done in 
\cite{Pelaez:2013cpa,Gracey:2019xom} for different gauge groups and renormalization schemes in the quenched case, and in \cite{Pelaez:2014mxa} including dynamical quarks. For the $SU(3)$ group and the 
IS scheme the initial conditions for the R-G flow at $\mu_0=1$ GeV obtained are the ones listed in table \ref{tab:condiciones_iniciales} .

\begin{table}[htbp]
\begin{tabular}{l|l|l|l|}
\cline{2-4}
                                        & $m_0$ (GeV) & $M_0$ (GeV) & $g_0$ \\ \hline
\multicolumn{1}{|l|}{Quenched}          & 0.35        & -           & 3.6   \\ \hline
\multicolumn{1}{|l|}{Unquenched ($N_f=2$)} & 0.42        & 0.13        & 4.5   \\ \hline
\end{tabular}
\caption{Values of the masses of the quark and gluon ($M_0$ and $m_0$ respectively) and the coupling constant ($g_0$) at renormalization scale $\mu_0=$1 GeV obtained by adjusting the 2-point functions to lattice data, 
both for the quenched case and the unquenched case with two degenerate quark flavors. }
\label{tab:condiciones_iniciales}
\end{table}

In this work we use these values to compute the one loop three-gluon vertex, which means that up to the overall normalization constant our results are a pure prediction of the model.

\subsection{Comparison with the lattice}

In order to compare with the lattice, we must choose specific momentum configurations for $\Gamma_{\mu \nu \rho}(p_1,p_2,p_3)$. Most available lattice data employs some of the following configurations: 
The \textit{Symmetric Configuration}, with $p_{1}^2=p_{2}^2=p_{3}^2=p^2$ and $p_1 \cdot p_2 = p_1 \cdot p_3 = p_2 \cdot p_3 = -\frac{p^2}{2}$, the \textit{Asymmetric Configuration}, with $p_1=0$ and $p_2=-p_3=p$, 
and the \textit{Orthogonal Configuration}, with $p_1 \cdot p_2 = 0$, $p_{1}^2=p_{2}^2=p^2$ and $p_{3}^2=2 p^2$. 

For the quenched case, we compared our results with the lattice data from \cite{Aguilar:2021lke}. Following their definitions, in the symmetric configuration we work with the scalar functions $ \bar{\Gamma}_{1}^{\mathrm{sym}}$ and $ \bar{\Gamma}_{2}^{\mathrm{sym}}$:

\begin{align}
\label{eq_defggammasym}
    g \Gamma_{ \mu \nu \rho}(p_1,p_2,p_3) ={}& \bar{\Gamma}_{1}^{\mathrm{sym}}\left(s^{2}\right) \lambda_{1}^{\mu \nu \rho}(p_1,p_2,p_3)\nonumber\\
         &+\bar{\Gamma}_{2}^{\mathrm{sym}}\left(s^{2}\right) \lambda_{2}^{\mu \nu \rho}(p_1,p_2,p_3),
\end{align}
where
\begin{align*}
 \lambda_{1}^{\mu \nu \rho}(p_1,p_2,p_3)&=\nonumber\\
 \Gamma_{\mu' \nu' \rho'}^{(0)}&(p_1,p_2,p_3) P^\perp_{\mu' \mu}(p_1) P^\perp_{\nu' \nu}(p_2) P^\perp_{\rho' \rho}(p_3),
\end{align*}
with $\Gamma_{\mu' \nu' \rho'}^{(0)}(p_1,p_2,p_3)$ defined as the perturbative tree-level tensor of the three-gluon vertex, and\\ $\lambda_{2}^{\mu \nu \rho}(p_1,p_2,p_3)=\frac{(p_1-p_2)_{\rho} (p_2-p_3)_{\mu} (p_3-p_1)_{\nu}}{p^2}$.

On the other hand the asymmetric configuration of the vertex is parametrized by a single scalar function $ \bar{\Gamma}_{3}^{\mathrm{asym}}$ defined by
\begin{equation}
g \Gamma_{ \mu \nu \rho}(p,-p,0)=\bar{\Gamma}_{3}^{a s y m}\left(p^{2}\right) \lambda_{3}^{ \mu \nu \rho}(p,-p, 0),
\end{equation}
with
\begin{equation}
\lambda_{3}^{ \mu \nu \rho}(p,-p, 0)=2 p^{\rho} P^{\perp\mu \nu}(p).
\end{equation}
We compared our unquenched results with the lattice data from \cite{Sternbeck:2017ntv}. 
They work in the orthogonal configuration, and define the usual scalar function $G_1$, which consists on contracting the external legs of the vertex with transverse propagators and the tree-level momentum 
structure of the 3-gluon vertex, normalized to the same expression at tree-level. This reads:
\begin{widetext}
\begin{equation}
\label{G1}
G_{1}(p,k,r)=\frac{\Gamma^\text{tree-level}_{\alpha \beta \gamma} (p,k,r)P^\perp_{\alpha\mu}(p)P^\perp_{ \beta\nu } 
(k)P^\perp_ { \gamma\rho } 
(r)\Gamma_{\mu\nu\rho}(p,k,r)}{\Gamma^\text{tree-level}_{\alpha \beta \gamma} (p,k,r)
P^\perp_{\alpha\mu}(p)P^\perp_{ \beta\nu } (k)P^\perp_ { \gamma\rho } 
(r)\Gamma^\text{tree-level}_{\mu \nu \rho} (p,k,r)}
\end{equation}
\end{widetext}
The results of the model are shown below using the different scalar functions defined in this section including the renormalization group effects.

\subsection{$SU(3)$ Yang Mills results}
We first present our results for $SU(3)$ Yang Mills theory and we compare them with lattice results from \cite{Aguilar:2021lke}. 
As stated before, we integrate the beta functions with initial conditions at $\mu_0=1$ GeV using the initial conditions listed in table \ref{tab:condiciones_iniciales}. 

\begin{figure}[htbp]
 \includegraphics[width=\linewidth]{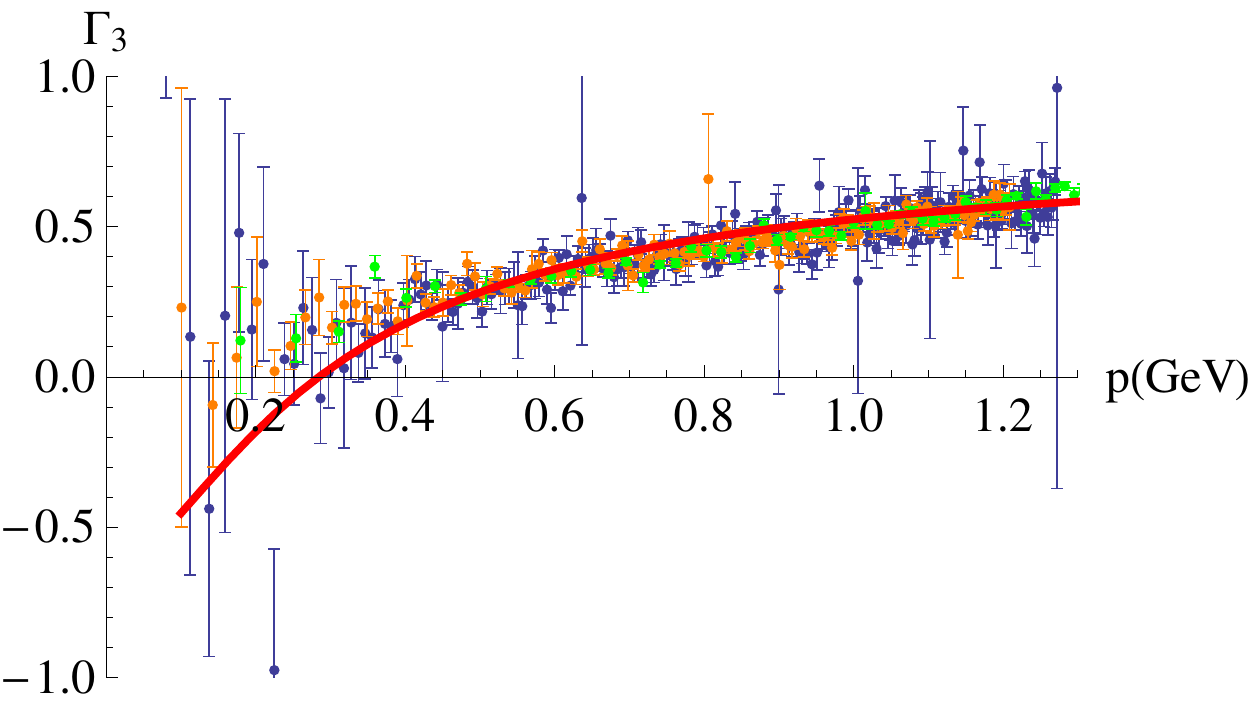}
\caption{$\bar{\Gamma}_{3}^{\mathrm{asym}}$ 
scalar function as a function of momentum for one vanishing momentum (asymmetric configuration). The points are lattice 
data from \cite{Aguilar:2021lke}. The plain red line corresponds to our 1-loop computation.}
\label{fig_asym}
 \end{figure}

\begin{figure}[htbp]
 \includegraphics[width=\linewidth]{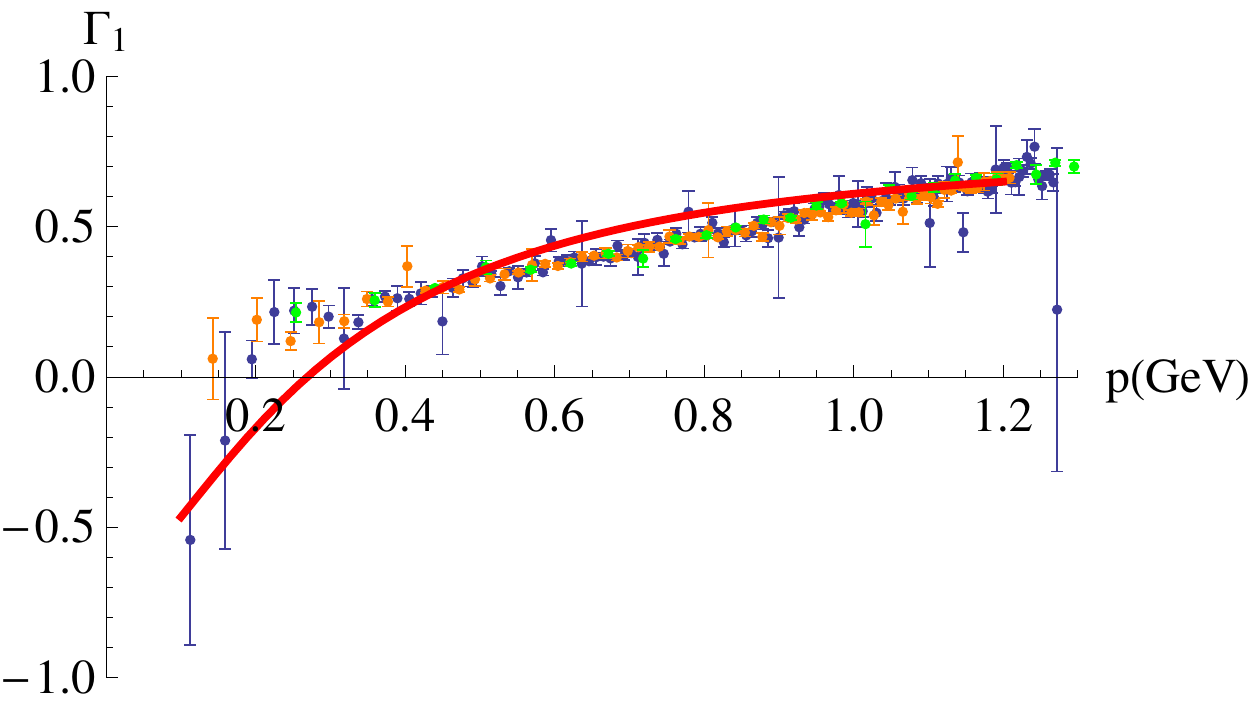}
 \includegraphics[width=\linewidth]{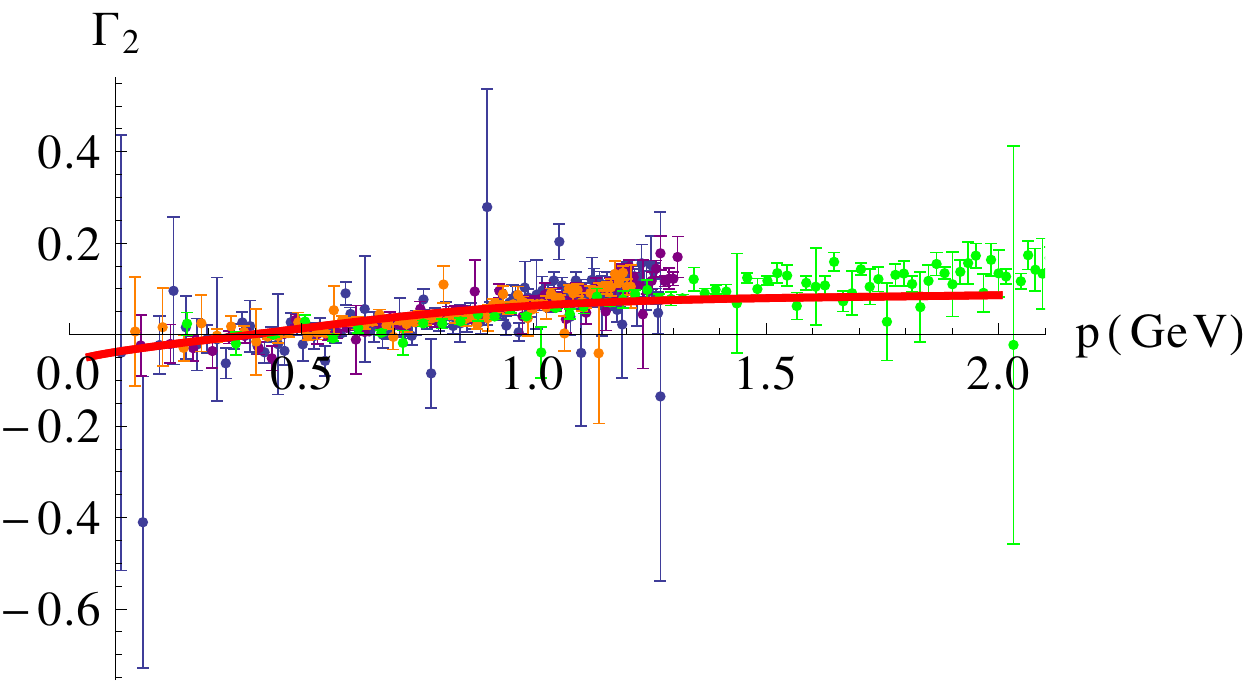}
\caption{$ \bar{\Gamma}_{1}^{\mathrm{sym}}$ (top) and  $ \bar{\Gamma}_{2}^{\mathrm{sym}}$
(bottom) scalar functions as a function of momentum for all momenta equal (symmetric configuration). The points are lattice 
data from \cite{Aguilar:2021lke}. The plain red line corresponds to our 1-loop computation.}
\label{fig_sym}
 \end{figure}

In Fig. \ref{fig_asym} we show the results for the scalar function $ \bar{\Gamma}_{3}^{\mathrm{asym}}$ in the asymmetric configuration (one vanishing momentum), 
and in Fig. \ref{fig_sym} we do the same for the functions  $ \bar{\Gamma}_{1}^{\mathrm{sym}}$ and  $ \bar{\Gamma}_{2}^{\mathrm{sym}}$ the symmetric configuration (all momenta equal). 

 In all cases the agreement is very good, 
 specially considering that the initial conditions for the renormalization-group flow were not fitted for the three-point function but for the propagators. 
 It is also noticeable that in all cases the different scalar functions become negative at low energies, 
 a qualitative feature that was observed in many lattice simulations as well as in different analysis \cite{Aguilar:2013vaa,Athenodorou:2016oyh,Boucaud:2017obn,Aguilar:2019jsj,Aguilar:2019uob,Aguilar:2019kxz}. 
 While the scalar functions associated to the tree-level tensor diverges logarithmically, the $\bar{\Gamma}_{2}^{\mathrm{sym}}$ goes to a constant value in the infrared as stated in \cite{Aguilar:2021lke}.
 The simplicity of one-loop CF model allows to write the infrared behaviour of $\bar{\Gamma}_{2}^{\mathrm{sym}}$ analytically:
 
\begin{align}
 \bar{\Gamma}_{2}^{\mathrm{sym}}\sim\frac{g^2 N}{414720 \pi ^2} \left(20 \left(16 \sqrt{3}
   \text{Cl}_2\left(\frac{\pi
   }{3}\right)-33\right) +189\frac{p^2}{m^2}\right),
\end{align}
which is indeed finite in the infrared, and where $\text{Cl}_2$ is the Clausen function satisfying $\text{Cl}_2\left(\frac{\pi}{3}\right)=1.0149417$. It is worth mentioning that this behaviour is not modified by the effects of the renormalization group.
 
 These results also show the divergent behavior of $\bar{\Gamma}_{1}$, which can be easily understood due to the presence of massless ghosts as stated in \cite{Pelaez:2013cpa}.

\subsection{Unquenched QCD results}

If we want to include the influence of dynamical quarks to the previous computation we must add the quark triangle diagram to the vertex and use the running of the coupling obtained 
in the unquenched analysis \cite{Pelaez:2014mxa}.
The contribution of that diagram can be computed with no difficulty in arbitrary dimension and  for arbitrary number of quarks. Our explicit expressions match the ones presented in \cite{Davydychev:2001uj}
when continuing them to Euclidean space. 
In order to be more specific we show as an example the explicit bare contribution of the quark-loop
diagram to the factor $G_1$ in $d=4-2\epsilon$ dimensions in Appendix \ref{Ap_UQ}. 

\begin{figure}[htbp]
 \includegraphics[width=\linewidth]{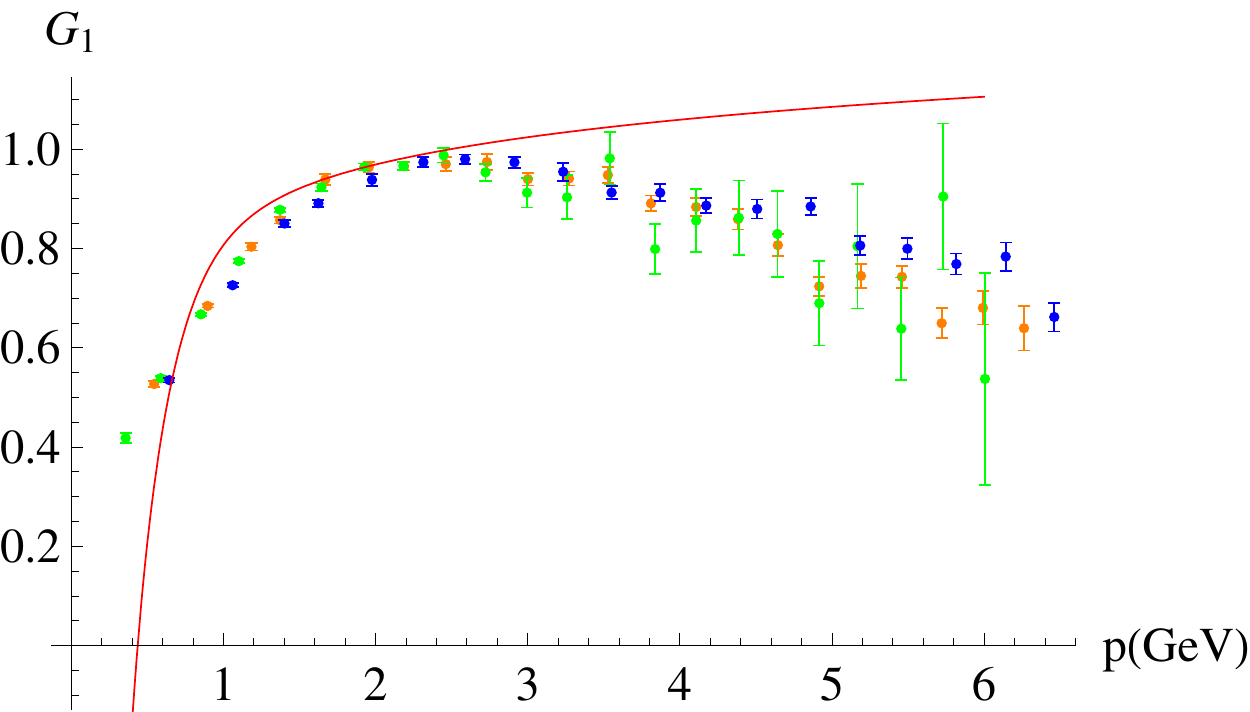}
 
\caption{$G_1$ scalar function as a function of momentum for two momenta orthogonal (orthogonal configuration) and two mass-degenerate quark flavors. The points are lattice 
data from \cite{Sternbeck:2017ntv}. The plain red line corresponds to our 1-loop computation.}
\label{fig_G1}
 \end{figure}

The total result for $G_1$ is shown in Fig.~\ref{fig_G1} where it is compared with lattice data from \cite{Sternbeck:2017ntv}.
The data available corresponds to the $G_1$ scalar function in the case of two mass-degenerate quark flavors ($N_f=2$) in the orthogonal configuration (two momenta orthogonal to each other and of equal magnitude).

In this case, the agreement is still very good in the infrared but worsens in the UV. 
More precisely, the model and the lattice results start separating at a scale of about 2.5 GeV. 
This scale is of the order of magnitude of the inverse of the lattice spacing used in most lattice simulations, 
and therefore lattice results beyond this scale are subject to hypercubic lattice spacing artefacts. 
Taking this fact into account and also considering that perturbation theory must work at one-loop in the UV, 
we suspected that the decrease in the values of $G_1$ after the inverse lattice spacing scale must be caused by finite lattice artefacts such lattice effects in the UV.

To confirm this statement, we computed analytically the high-energy limit of $G_1$, finding that it behaves in the UV as $\text{Ln}(\frac{p}{\mu_0})^{\alpha}$ with $\alpha=\frac{17N-16 N_f T_f}{44N-16 N_f T_f}$, 
which is compatible with our results. 
The idea behind this computation is that since the UV limit of $G_1(p,p)$ is equal to 1, the high-energy behavior of $G_1(\mu_0,p)$ must be given by $z_{A}^{-\frac{3}{2}}$ 
as a consequence of the renormalization-group equation given in Eq. (\ref{eq_RGequation}). 
The full computation show that the UV limit of $z_{A}^{-\frac{3}{2}}$ is indeed $z_{A}^{-\frac{3}{2}} \propto \text{Ln} (\frac{\mu}{\mu_0})^{\frac{35}{116}}$ for $N=3$ and $N_f = 2$. 
In conclusion, our one-loop computation matches the lattice results in their regime of validity and the renormalization-group prediction in the UV.
 
\begin{figure}[htbp]
 \includegraphics[width=\linewidth]{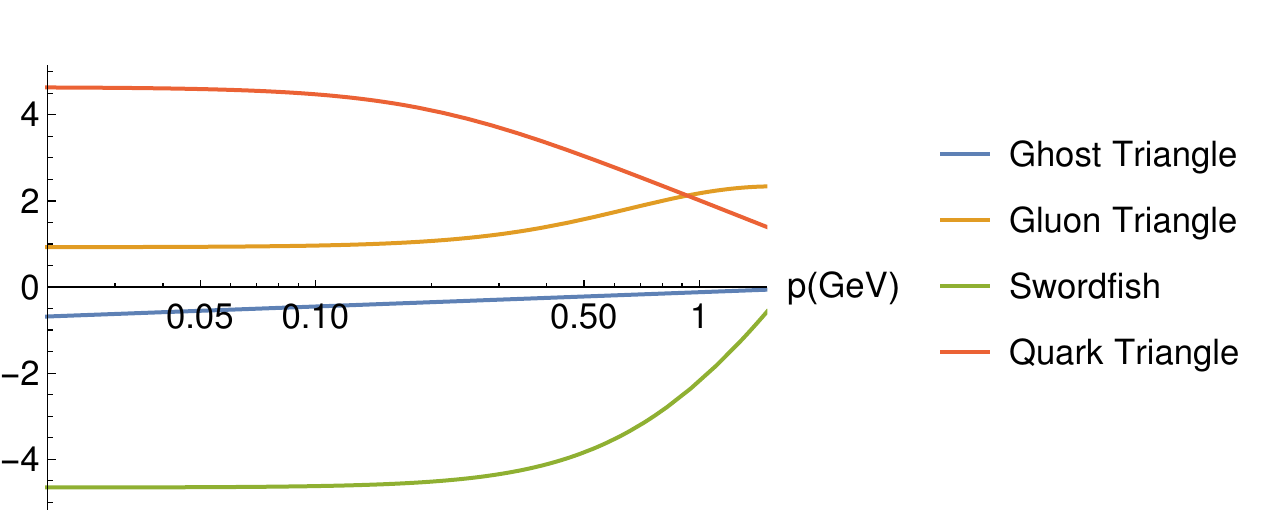}
\caption{\label{fig_diagramas}Infrared behaviour of one-loop diagrams contributing to the three-gluon vertex in logarithmic scale. The ghost triangle contribution goes as $\log p$ in the IR while the other Feynman diagrams go to a finite value.}
 \label{fig_contribucionesalvertice}
\end{figure}

To complete the analysis we include the contributions to the three-gluon vertex arising from the different one-loop Feynman diagrams, which are shown in the asymmetric configuration in Fig.~\ref{fig_diagramas}. Our results match the behaviour observed in \cite{Huber:2016tvc} for the Yang-Mills contributions, and indeed show that the divergence of the vertex in the infrared is caused by the ghost triangle's logarithmic divergence.

\subsubsection{Zero crossing and the number of flavours.}

 In this section we study the influence of dynamical quarks in the position of the zero crossing of the three-gluon vertex. 
 In the one-loop CF-model the influence of quarks in the renormalized vertex can be isolated as the term proportional to $N_f$. 
 At one loop, the quarks contribution to the renormalized vertex is proportional to:
 
 \[\mathcal{F}=\left(\frac{3}{2}\delta Z_A^{UQ}\bigg|_{finite}-\delta Z_g^{UQ}\bigg|_{finite}+G_1^{UQ}\bigg|_{finite}\right)\]
 where $\delta Z_A^{UQ}\bigg|_{finite}$, $\delta Z_g^{UQ}\bigg|_{finite}$ and $G_1^{UQ}\bigg|_{finite}$ represent the finite part of the coefficient proportional to $g^2 N_f T_f$ in the $Z_A$- and $Z_g$-renormalization factor and in $G_1$  respectively.
 
   \begin{figure}[h!]
 \includegraphics[width=\linewidth]{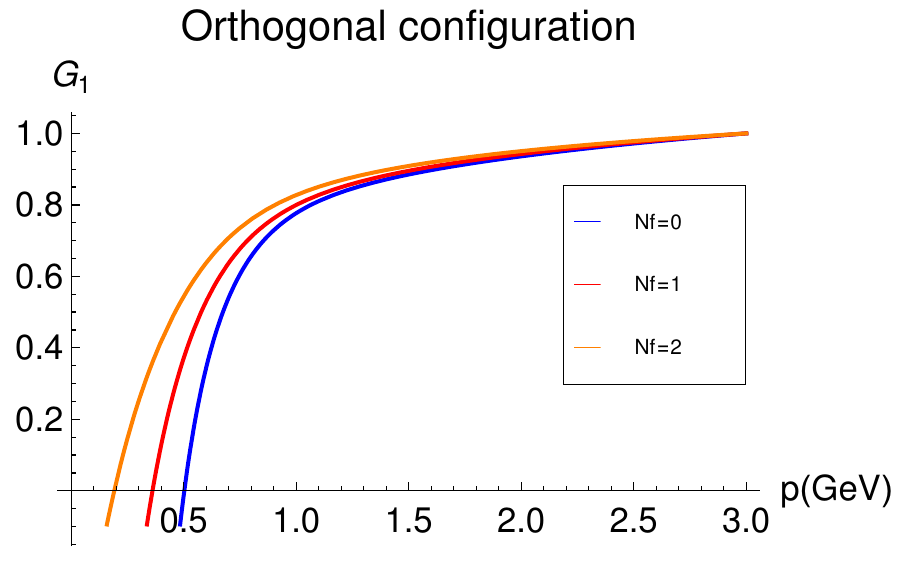}
  \includegraphics[width=\linewidth]{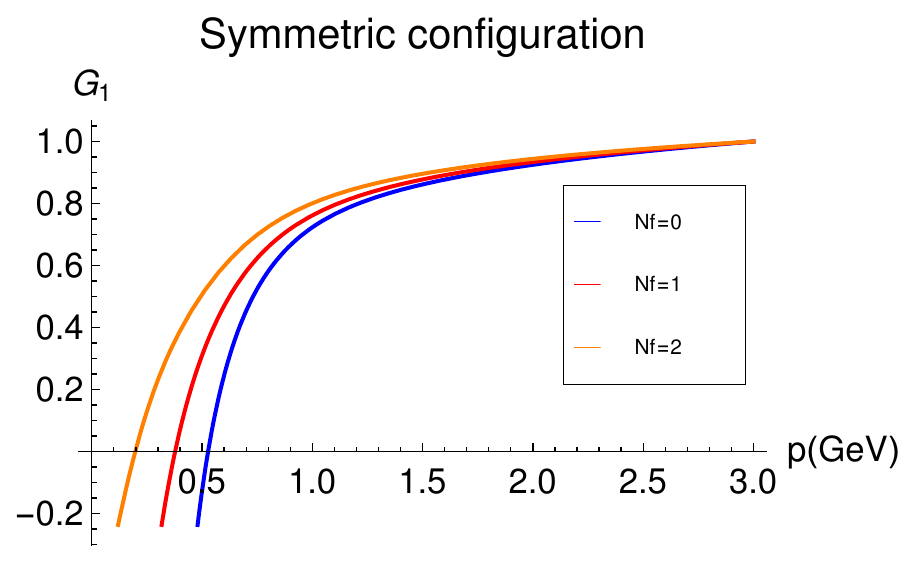}
   \includegraphics[width=\linewidth]{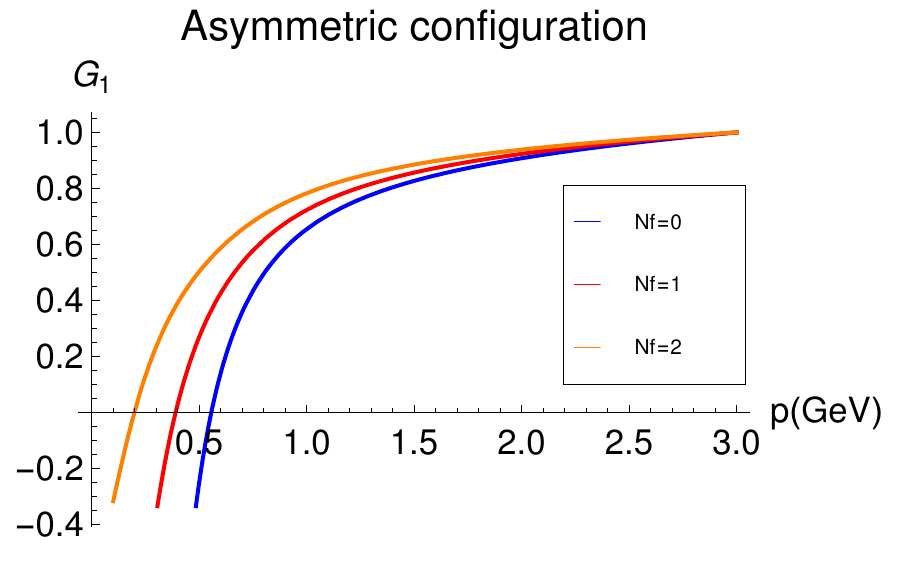}
   \caption{\label{fig_aaaQvsUQ} Comparison of the three-gluon vertex for different values of $N_f$ in the orthogonal configuration (top), symmetric configuration (middle) and gluon vanishing momentum (bottom)}
 \end{figure}

  We studied the sign of the factor $\mathcal{F}$ in order to analyze in which direction the zero crossing is shifted.  As $\mathcal{F}$ depends on the finite part of the renormalization factor, it is expected that its sign depends on the chosen renormalization scheme. For the IS-scheme we observe that even though the contribution of the quark triangle diagram is positive (see Fig.\ref{fig_diagramas}) the renormalization factors together with the renormalization group flow makes $\mathcal{F}$ negative. This means that the whole contribution arising from considering dynamical quarks is negative when compared to the quenched quantity using the same flow. However, as flows should be different in each situation, it is better to study the influence of dynamical quarks in the infrared region using the same set of initial conditions of the renormalization group flow at an ultraviolet scale (for instance $\mu=3$ GeV). In this context, we can see in Fig.~\ref{fig_aaaQvsUQ} that dynamical quarks shift the zero crossing to the infrared as it is also observed by \cite{Williams:2015cvx,Blum:2016fib}.


 \subsubsection{Unquenching the ghost-gluon vertex.}

  \begin{figure}[h!]
 \includegraphics[width=\linewidth]{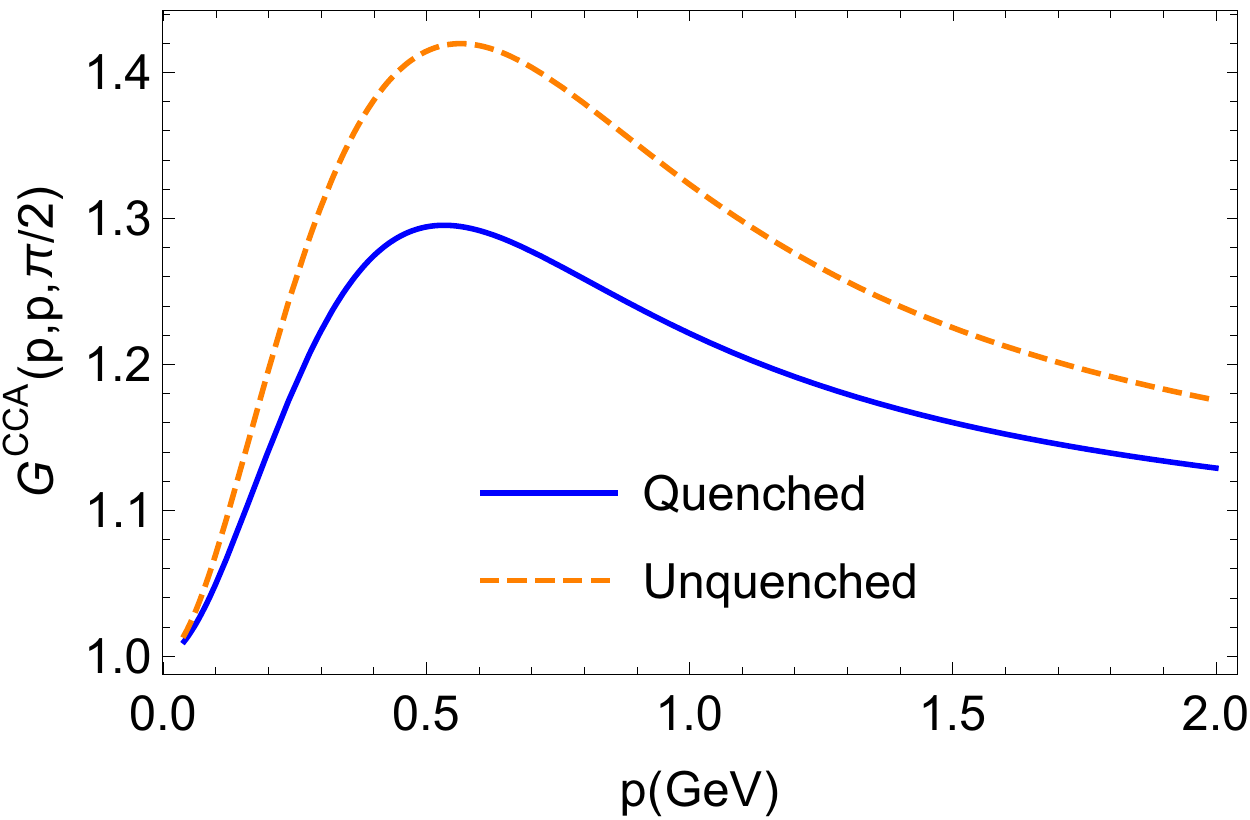}
  \includegraphics[width=\linewidth]{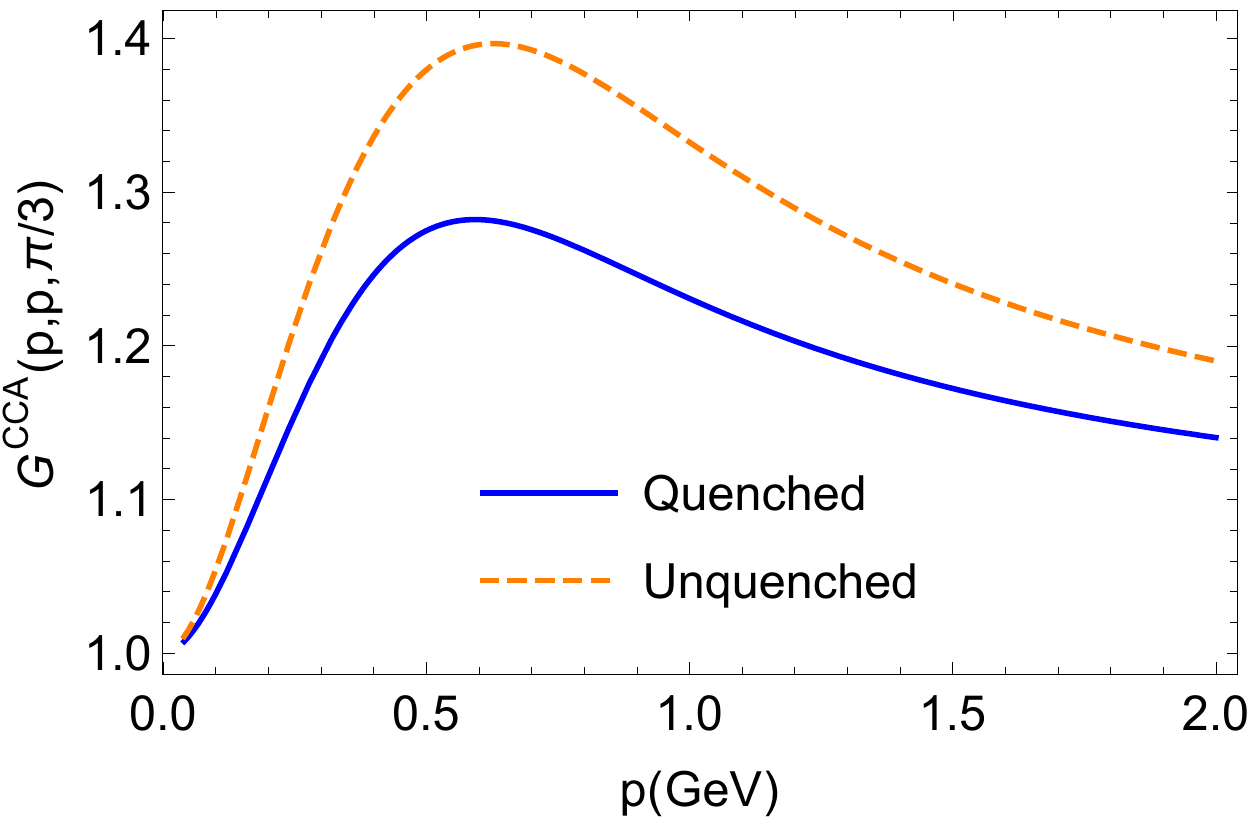}
   \includegraphics[width=\linewidth]{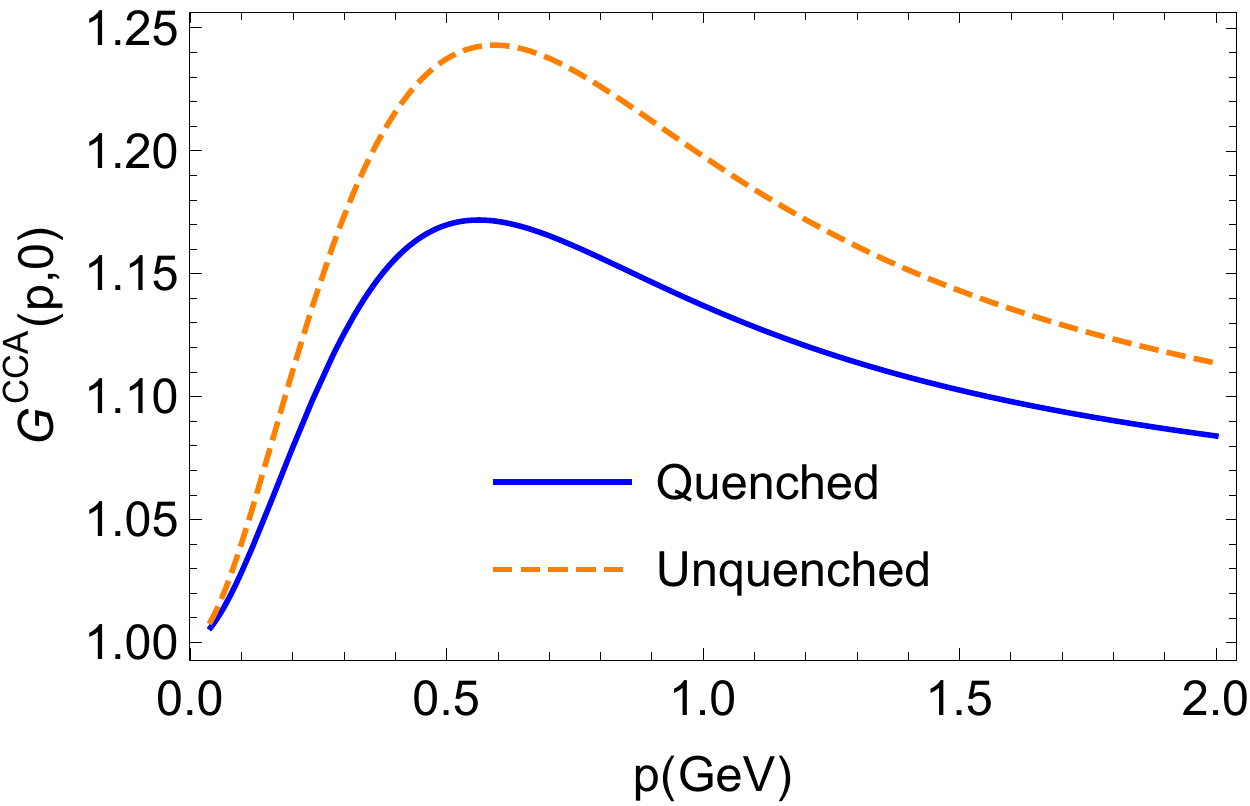}
   \caption{\label{fig_ccaQvsUQ} Comparison of the ghost-gluon vertex in the quenched and unquenched case using the set of parameters of Table~\ref{tab:condiciones_iniciales} in the orthogonal configuration (top), symmetric configuration (middle) and gluon vanishing momentum (bottom)}
 \end{figure}

   \begin{figure}[htbp]
  \includegraphics[width=\linewidth]{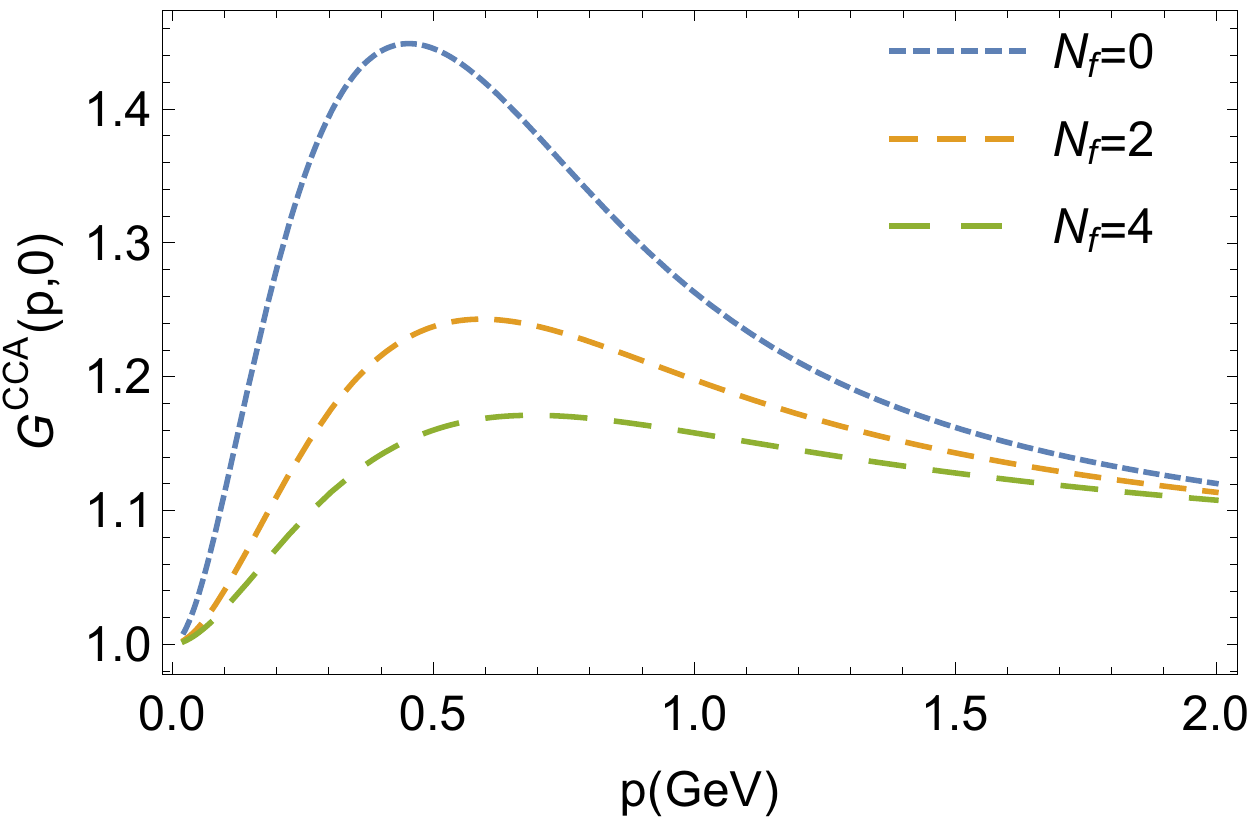}
 \caption{\label{fig_ccAcambioNF}Comparison of the ghost-gluon vertex varying $N_f$ using the same set of parameters at 3 GeV.}
 \end{figure}
 
   \begin{figure}[htbp]
 \includegraphics[width=\linewidth]{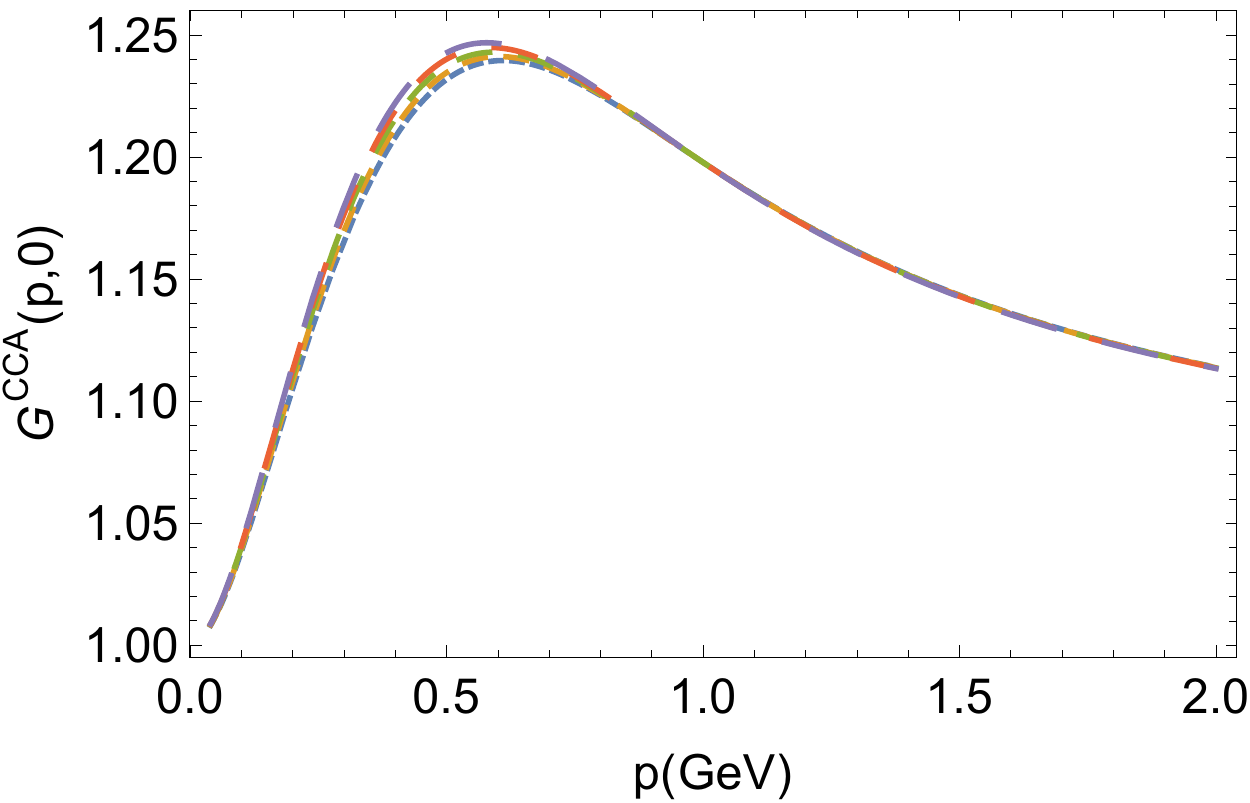}
 \caption{\label{fig_ccAMasas}Ghost-gluon vertex for different values of $M_0$ with fixed $g_0$ and $m_0$.}
 \end{figure}
 
 It is also interesting to observe the influence of dynamical quarks in the ghost-gluon vertex. Eventhough
 quarks do not contribute directly in its one-loop diagrams the inclusion of dynamical quarks affects its renormalization group flow.
 
 The function $G^{c\cb A}(p,k,r)$ defined through the vertex as:
 \begin{equation}
\label{GccbA}
 G^{c\cb A}(p,k,r)=\frac{k_\nu P^\perp_{\mu\nu}(r)
\Gamma_{\mu}(p,k,r)} {k_\nu P^\perp_{\mu\nu}(r)k_\mu }
\end{equation}
 where
  \begin{equation}
  \label{eq_vertex_cbcA_def}
  \Gamma^{(3)}_{c^a\cb^b A_\mu^c}(p,k,r)=-i g_0 f^{abc} \Gamma_{\mu}(p,k,r)
\end{equation}
is shown in Fig.~\ref{fig_ccaQvsUQ} for different kinematical configurations using the parameters of Table~\ref{tab:condiciones_iniciales} which are the parameters that give a better fit to one-loop propagators. It is important to mention that $G^{c\cb A}(p,k,r)$ is renormalized by the combination $Z_g\sqrt{Z_A}Z_c$ which is set to one accordingly to Taylor scheme (\ref{eq_taylor}).  
Therefore the value of ghost-gluon function only depends on the values of $g$ and $m$ on each momentum scale. In particular, initial conditions of the renormalization flow are initialized at $1$GeV, therefore the vertex function at that scale is purely determined by the values from Table~\ref{tab:condiciones_iniciales}. As the value of $g_0$ is larger in the unquenched case, the unquenched vertex function will be above the quenched one as is observed in Fig.~\ref{fig_ccaQvsUQ}.

To study the influence of dynamical quarks on the ghost-gluon vertex is therefore convenient to raise the flavor number but using the same initial conditions of the flow. In Fig.~\ref{fig_ccAcambioNF} we compare the ghost-gluon vertex when raising $N_f$ using the same set of parameters at $3$GeV. It can be seen that the unquenching effects reduces the vertex contribution in the infrared. Another observation is that the unquenched ghost-gluon vertex is almost insensitive to the value of the quark mass as it is seen in Fig.~\ref{fig_ccAMasas}.

\section{Conclusions}
\label{sect_conclusion}
With the aim of studying the infrared properties of the gluon and ghost-gluon three-point correlation functions we presented a one-loop
calculation using Curci-Ferrari model in Landau
gauge for arbitrary kinematical configurations. 
The results are an extension of a previous work \cite{Pelaez:2013cpa} to the unquenched case.
In particular, we compared the results for the vertex with the available lattice data including dynamical quarks corresponding
to the kinematical configuration with a vanishing gluon momentum and two degenerate flavors.
A study of the position of the zero crossing of the vertex was also done, observing that the position of the zero crossing is shifted to the infrared due to the presence of dynamical quarks when compared to the quenched case.

We also studied the quenched case because some infrared properties observed by the model in the previous work 
were not clear in lattice simulations at that time. However, the infrared lattice study of the three-gluon 
vertex has improved in the last years and now error bars are good enough to understand its 
infrared behavior as discussed in \cite{Aguilar:2021lke}. 
In particular, lattice simulations show a change of sign in the deep infrared that is easily 
understood by the Curci-Ferrari model. 
As it has been discussed in \cite{Pelaez:2013cpa,Aguilar:2013vaa} it can be explained as a 
consequence of the diagram with a loop of massless ghosts. 
In the quenched case we compare the results with lattice simulations for the completely symmetric and 
the antisymmetric configuration. 
The results show an excellent match to lattice results, specially considering that the free parameters of the model were already 
fixed by fitting the propagators.

\begin{acknowledgments}

The authors would like to acknowledge the financial support from PEDECIBA and 
ECOS program and from the ANII-FCE-126412 project. We also thank N. Barrios, U. Reinosa, M. Tissier and N. Wschebor for useful discussions.
\end{acknowledgments}

\appendix

\section{Contribution of dynamical quarks}
\label{Ap_UQ}
The contribution of dynamical quarks to the $G_1$ factor:
\begin{widetext}
\begin{equation}
\label{factorGunquenched}
\begin{split}
&\frac{g^2 N_f T_f}{12 \pi ^2 \epsilon}+
\frac{g^2 N_f T_f }{12 \pi ^2 }\bigg[-\frac{1}{3}+\frac{2 \chi\left(A_0\left(M^2\right)+M^2\right)}{D_1}+\frac{1}{D_1 D_2}
\bigg(24 k^2 p^2 r^2 C_0\left(M^2,M^2,M^2,p,r\right) \left(\mathcal{S}-M^2 \chi\right)\\
   &+\mathcal{K}_1 B_0\left(M^2,M^2,-p^2\right)+\mathcal{K}_2 B_0\left(M^2,M^2,-r^2\right)+\mathcal{K}_3   B_0\left(M^2,M^2,-k^2\right)\bigg)\bigg]
\end{split}
\end{equation}
\end{widetext}  
where $A_0$, $B_0$ and $C_0$ are the corresponding finite part of master integrals defined in Eq.~(\ref{masters}) and
\begin{equation*}
\begin{split}
&\chi=p^2+r^2+k^2,\\
&\mathcal{S}=p^2r^2+p^2k^2+r^2k^2,\\
&D_1=k^4+p^4+r^4+10\mathcal{S},\\
&D_2=k^4+p^4+r^4-2\mathcal{S}
\end{split}
\end{equation*}
and
\begin{equation*}
\begin{split}
\mathcal{K}_1&=k^6 \left(2 M^2-p^2\right)+k^4 \left(2 M^2 \left(9 p^2-r^2\right)-9 p^4-11 p^2 r^2\right)\\
&-k^2 \left(2M^2 \left(9 p^4+2 p^2 r^2+r^4\right)-9 p^6+10 p^4 r^2+11 p^2 r^4\right)\\
&-\left(2 M^2-p^2\right)
   \left(p^6+9 p^4 r^2-9 p^2 r^4-r^6\right)
\end{split}
\end{equation*}
while $\mathcal{K}_2$ and $\mathcal{K}_3$ are obtained exchanging $p\leftrightarrow r$ and  $p\leftrightarrow k$ in $\mathcal{K}_1$ respectevely.

\end{document}